# 15µm Quantum well infrared photodetector for thermometric imagery in cryogenic windtunnel

Emmanuel Lhuillier[a,b], Isabelle Ribet-Mohamed[a], Nicolas Péré-Laperne[a], Michel Tauvy[a], Joël Deschamps[a], Alexandu Nedelcu[c], Emmanuel Rosencher[a]
[a]*ONERA, centre de Palaiseau, Chemin de la Hunière- FR 91761 Palaiseau cedex, France.*
[b]*Laboratoire Matériaux et Phénomènes Quantiques, Université Paris Diderot, CNRS UMR 7162, Bâtiment Condorcet, Case 7021, 75205 Paris cedex 13, France.*
[c]*Alcatel-Thales III-V Lab, Campus de Polytechnique, 1 Avenue A. Fresnel, 91761 Palaiseau cedex, France.*

**Abstract:**
Quantum Well Infrared Photodetector (QWIP) usually suffer from a too moderate quantum efficiency and too large dark current which is often announced as crippling for low flux applications. Despite this reputation we demonstrate the ability of QWIP for the low infrared photon flux detection. We present the characterization of a state of the art 14.5µm QWIP from Alcatel-Thales 3-5 Lab. We developed a predictive model of the performance of an infrared instrument for a given application. The considered scene is a Cryogenic Wind Tunnel (ETW), where a specific Si:Ga camera is currently used. Using this simulation tool we demonstrate the QWIP ability to image a low temperature scene in this scenario. QWIP detector is able to operate at 30K with a NETD as low as 130mK. In comparison to the current detector, the temperature of use is three times higher and the use of a QWIP based camera would allow a huge simplification of the optical part.

*PACS*: 85.60Bt, 85.60Gz

*Keyword*: Infrared photodetector, Low flux scenario.

## 1. Introduction

According to the Wien law, high wavelength detectors are good candidates for low photon flux detection which is involved in applications such as defence [1], astronomy [2] or cryogenic wind tunnels [3]. QWIPs are well-known to provide high uniformity focal plane arrays (FPAs) with few bad pixels. Moreover band gap engineering allows to reach any desired wavelength between 3 and 70µm. However their relatively low quantum efficiency and high dark current are often pointed out as severe drawbacks. Thus, in spite of many reports about the use of QWIP for high wavelength and low flux application [1,4,5], QWIP still suffers from a bad reputation in comparison with the MCT device. In this paper we demonstrate the ability of a 14.5µm QWIP to image of a low temperature scene (120K).

We first present the design of a 14.5µm QWIP dedicated to the low flux detection. Then, in section 3, we present the characterization of this detector under low infrared photon flux. Our bench allows the illumination of our detector with a flux which varies over nearly four orders of magnitude: $10^{12}/10^{16}$ photons·s$^{-1}$·cm$^{-2}$. We underline the good linearity of the component over this range of incident power. Section 4 is dedicated to the prediction of the performance of a camera based on this QWIP detector. We use an academic low flux scenario which is the cryogenic wind tunnel. We predict that this QWIP would succeed in imaging a scene with a temperature as low as 120K with a NETD of 130mK. The way on which improvement has to be focused is also suggested.

## 2. Sample

The considered QWIP is a forty period structure GaAs/$Al_{15.2}Ga_{84.8}As$. The well (barrier) width is 7.3nm (35nm). Each well is Si doped in its central part with a sheet concentration of $3\times10^{11}$cm$^{-2}$. The QWIP is processed into mesa of 23.5µm length and a grating is etched on the surface of each pixel. The structure is much more described in ref [6,7].

## 3. Characterization

We performed a full characterization of the component (I(V), noise, differential resistance and quantum efficiency) under low flux by changing the applied bias, detector temperature and black body temperature.

Dark current measurements revealed the existence of a plateau [6] in the I(V) curves in the tunnelling regime which is attributed to the competition between the electric field localisation and the barrier lowering by the electric field [7,8]. Results are given in the last section of this paper. Dark current measurements will not be further discussed in this paper but the plateau can clearly be seen in FIG. 1. The spectral responsivity is peaked at 14.5µm with a full half band width of 2µm (see inset of FIG. 1).

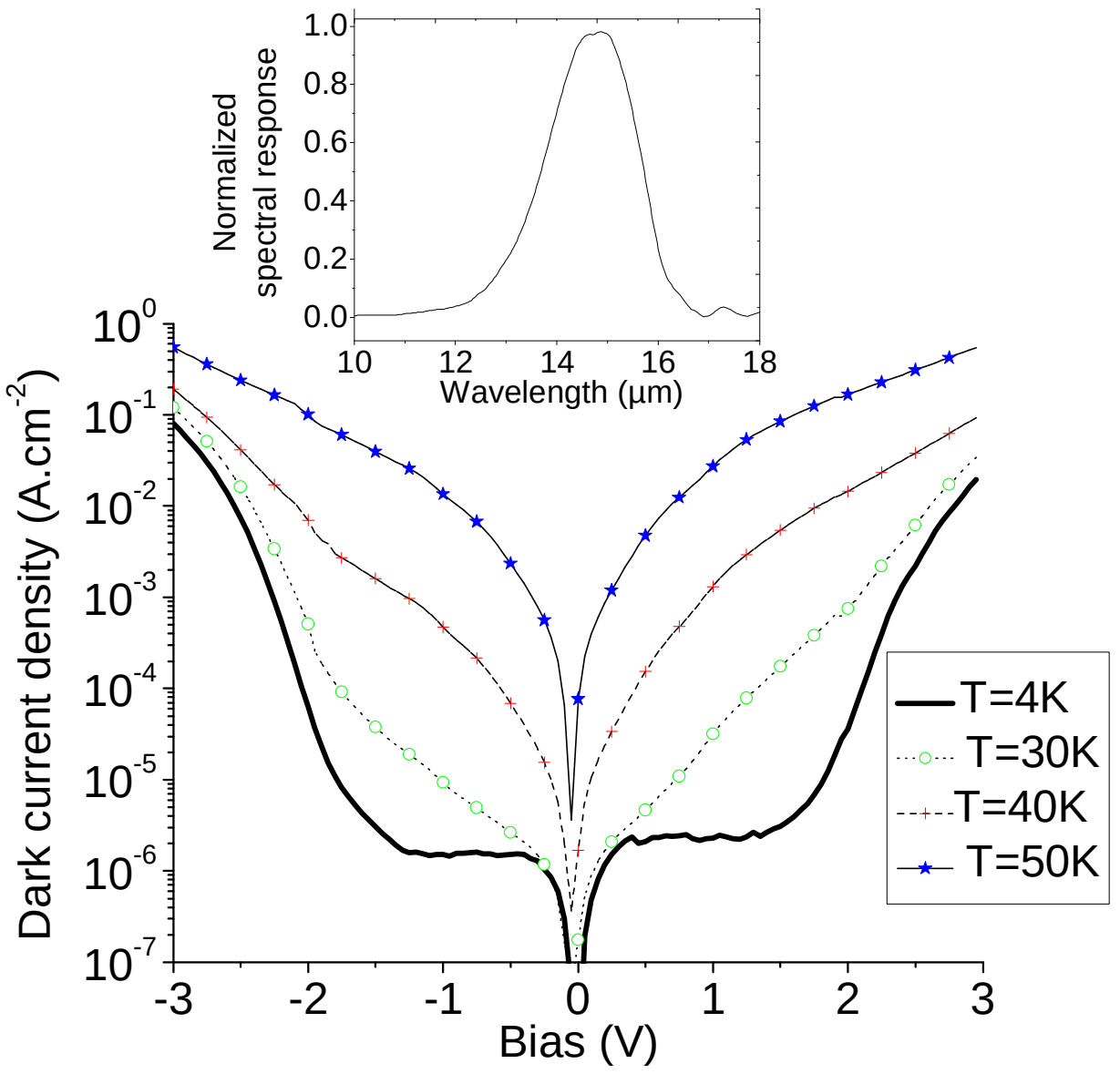

FIG. 1 Dark current as a function of the applied bias for different temperatures. Inset Normalized spectral response under -1V and T=4K.

In this paper we focus on responsivity measurements with emphasis on the linearity of this response. The good linearity of the QWIP has already been reported [9,10,11] up to six decades, but authors are generally interested in higher photon flux regime of illumination (1µW·cm$^{-2}$ to 0.1GW·cm$^{-2}$). Many papers dealing with the linearity of detector over several orders of magnitude of flux, use a laser as an infrared source [11], we prefer a blackbody which is expected to be closer to the target spectrum. To reach the expected flux on the detector, around $10^{13}/10^{14}$ photons·s$^{-1}$·cm$^{-2}$, we used a cryogenic blackbody. Thus we developed a specific bench for the illumination of our detector. This bench is based on two cryostats. The first one, for the detector, is cooled with liquid helium whereas the second one, for the blackbody, is cooled with liquid nitrogen. Specific care was used to design entangled cold shields to get rid of stray light. The numerical aperture of the bench is 2.8.

By adjusting the nitrogen flow inside the second cryostat we can adjust the blackbody temperature between 77K and 340K. In practice we have explored temperatures between 90K and 270K. This bench allows us to explore more than three orders of magnitude, typically between 20nW·cm$^{-2}$ and 0.12mW·cm$^{-2}$ ($1.5\times10^{12}$photons·s$^{-1}$·cm$^{-2}$ and $8.4\times10^{15}$photons·s$^{-1}$·cm$^{-2}$).

The FIG. 2 shows the I(V) curves measured for different temperatures of the blackbody. From 80K for the blackbody temperature we can discern the photocurrent from the dark current. We observed that for a bias above 2V the dark current starts to be dominant. In order not to be dark current limited, the operating point of the QWIP will be on the $|V| < 2V$ range.

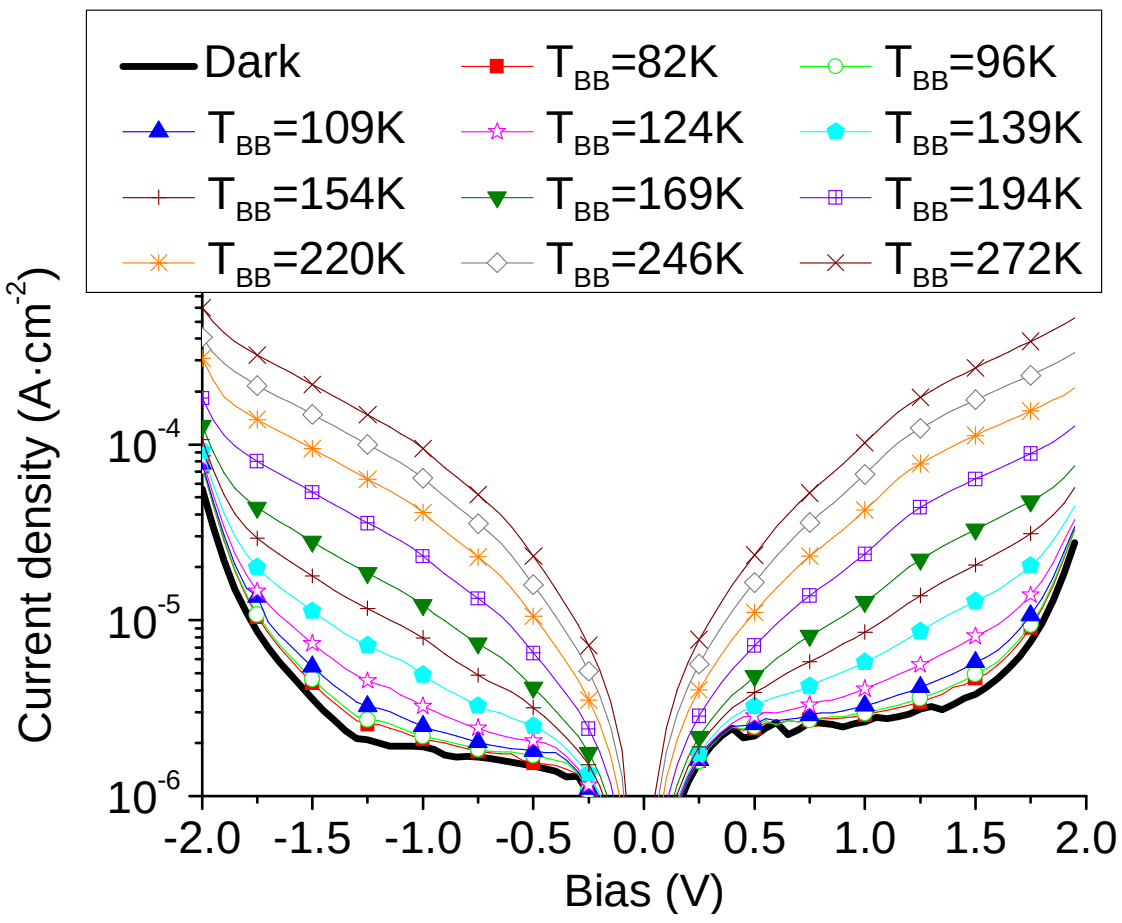

FIG. 2 I(V) curves under infrared photon flux for a blackbody temperature between 82K and 272K and a 4K detector temperature.

From the previous graph (FIG. 2) it is possible to extract the external quantum efficiency $\eta g$, which we define as the derivative of the total current density with respect to the flux divided by the proton charge. We assume that the current is locally linear as a function of the flux and thus we calculate the external quantum efficiency as:

$$\eta g = \frac{J_{total}(\phi_1) - J_{total}(\phi_2)}{e(\phi_1 - \phi_2)} \qquad (1)$$

where $\phi_1$ and $\phi_2$ are two different fluxes and $J_{total}(\phi)$ the current density associated with the $\phi$ flux.

Over more than three decades of flux the external quantum efficiency is almost constant, see FIG. 3. An external quantum efficiency of 7% is measured under -1V and 20% under -1.6V. Such values are in the top of the range of the reported values for VLWIR QWIP [12]. Above 2V dark current prevails, phenomenons such as impact ionisation [13] may occur, the noise is also important and those regimes of transport are not an interesting operating point.

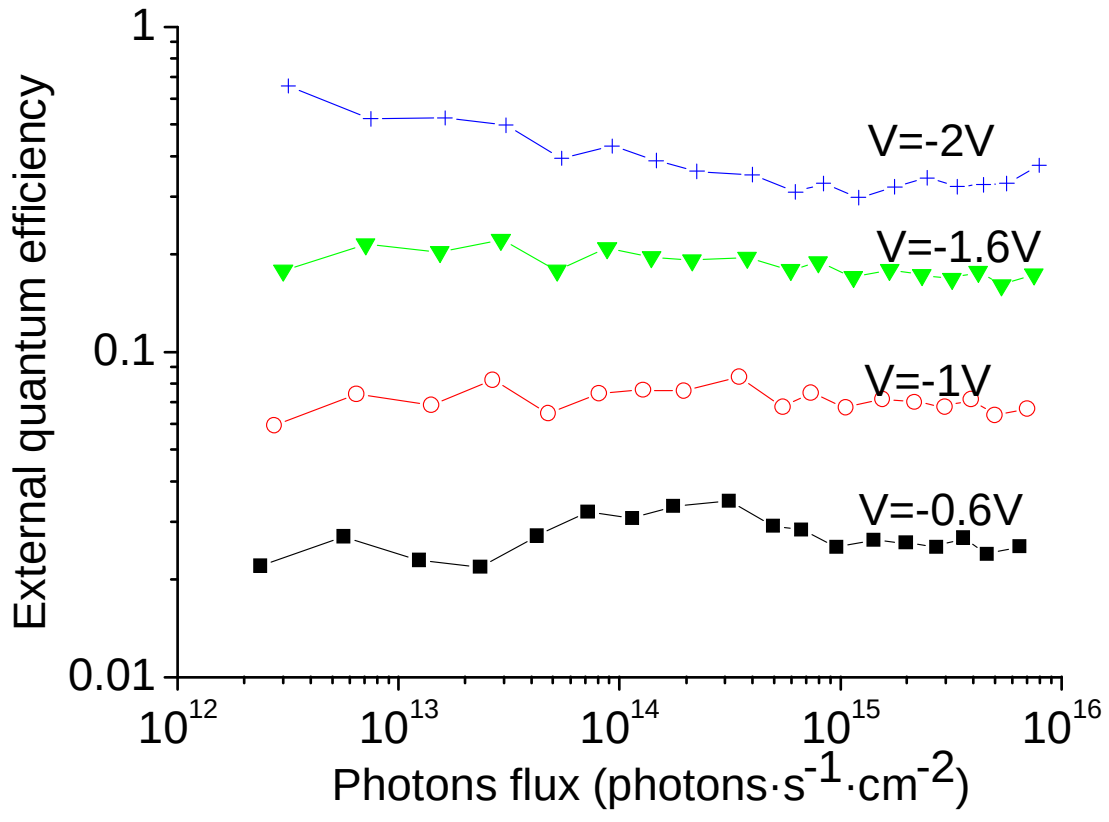

FIG. 3 External quantum efficiency as a function of the photon flux for different applied bias.

From our complete electro-optical characterisation, we know that our 14.5µm QWIP reaches the state of the art performances in terms of dark current and quantum efficiency. However those parameters do not allow us to conclude easily on the potential of our QWIP detector for low flux applications. To answer this question a system approach is required and presented in the next section.

## 4. Performance prediction for a low temperature scene

To predict the performance of an instrument we need a simulation tool. ONERA have developed a simulation code dedicated to the scaling of high performance infrared instrument. This home made code is fully analytic which make it very fast. This code has been validated by comparing its prediction with other simulation tool. It takes into account a target, an atmosphere, an instrument. The infrared FPA is an element of this instrument. The group instrument, atmosphere and target set up a scenario. This code is a proportioning tool, with a typical accuracy of ten percent.

Different scenarios are generally mentioned when dealing with low flux applications. Generally they are linked with defence or astronomy [1,14,15,16]. We have chosen an alternative way, which is thermometric imagery for the ETW (European transonic wind tunnel) cryogenic wind tunnel [3]. Nevertheless our result can easily be transposed to the previously evoked applications. The choice of such a scenario is allowed by the fact that ONERA has realized a dedicated camera [17] for this wind tunnel and thus all the associated parameters of the scenario are known.

### 4.1. ETW scenario and Crystal camera

The ETW cryogenic wind tunnel was built in Germany in the 90's in order to reach high Reynolds number ($Re>10^7$) which are particularly important to simulate the take off and landing of the airplanes. In order to increase the Reynolds number of the flow, it was decided to decrease drastically its temperature. For this purpose the air in the wind tunnel is cooled with liquid nitrogen, thus the temperature of the flow varies between 100K and 300K. The target is a wing model on which transition triggering device has been laid down to generate a forced laminar-to-turbulent transition. Such a transition is accompanied by a 1K change of the temperature which has to be detected by the camera. The tab. I sums up the required parameters of the infrared imaging device of the ETW.

ONERA was appointed by ETW to build an infrared camera which was able to detect a difference of temperature of 1K over a 120K scene. If we go back to the Wien law, the low temperature scene means that the maximum of the radiance of the model is in the mid IR range of wavelength (10µm to 30µm). Thus our laboratory had to design a specific camera [17]: Crystal. This camera was based on a Si:Ga FPA. Each pixel is 75µm large. The operating temperature is around 10K, with a spectral response included in 5-17µm range. Because of the small Focal Plane Array (FPA) size (192×128 pixels), a unique optic can not simultaneously answer to the resolution and field of view needs. Thus two optical blocks were required with two different focal lengths. On each optical path, it was possible to insert a density of 10 (required to avoid saturation of the detector when the scene temperature increase. All these optical elements had to be cooled to reduce the background radiation, which results in high requirements for the cooling system.

A value of 100ms was chosen for the integration time, this value is compatible with the typical timescale of the observed phenomenons.

tab. I Required performances for the infrared camera imaging the ETW scene

| Quantity | Value | Comment |
|---|---|---|
| Background temperature | 120K | |
| Model temperature | 121K | |
| Maximun linear range | 140K | Gives the maximum |
| temperature | | integration time. |
| Required NETD | <0.2K | |
| Integration time | ≤20 ms | Given by the gas flow dynamics |
| Required field of view | 40 cm | Model size |
| required spatial resolution | 2mm | 1 cm corresponds to the thermal marker size on the model |
| Distance from camera to model | 1 m | |

This Si:Ga camera succeeds in imaging this cryogenic scene, see FIG. 4.

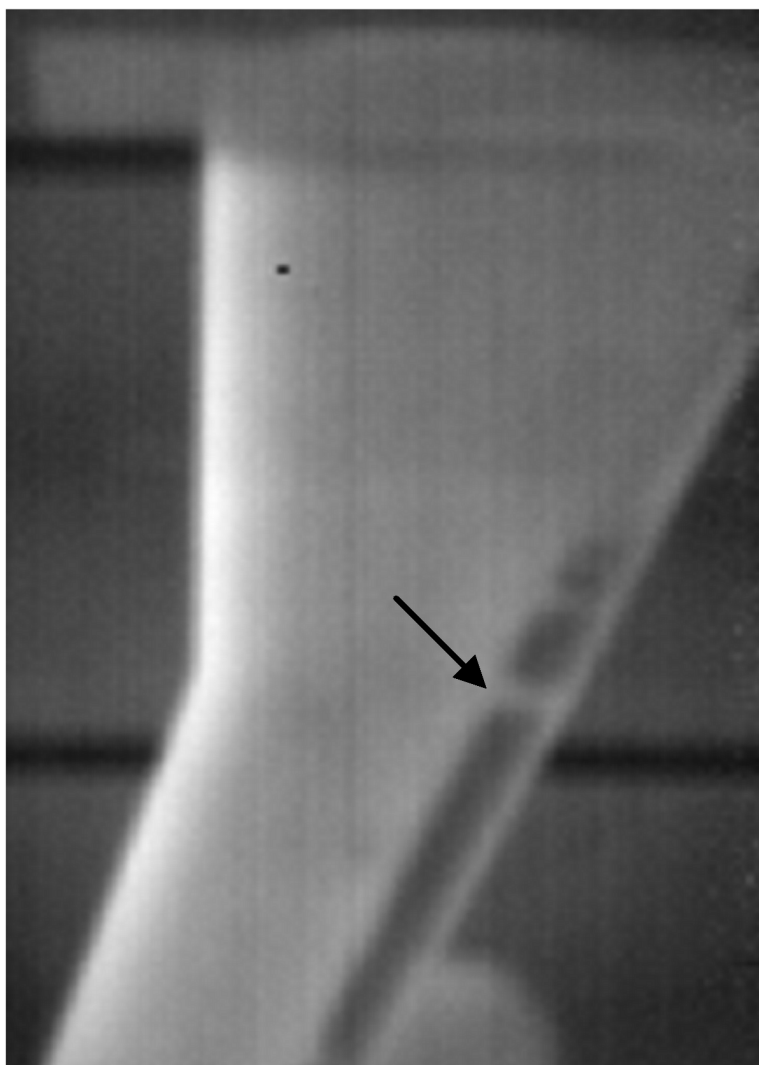

FIG. 4 Picture of wing model by the Crystal camera. The arrow shows the position of the transition triggering device, its temperature is 1K higher than the neighbouring part of the wing.

### 4.2. Camera based on the 15µm QWIP

In spite of the success of the camera Crystal to realize picture into the ETW wind tunnel, the low detector temperature and the complexity of the optical block plead in favour of the design of a new camera. In particularly we want to test the possibility to exchange this Si:Ga detector by one based on the described QWIP.

We have chosen to focus our interest on the operating temperature of the detector. As a consequence we operate this QWIP under a bias which gives its best performance. We have already discussed that the bias operating point is in the $|V| < 2V$ range. We will demonstrate in the following that the dark current is the limiting point of this detector. As a consequence we operate this QWIP near to the bias which maximizes the ratio of the total current over the dark current, see FIG. 5. This ratio reaches a maximum for bias around 1.4V with a small asymmetry with the bias polarity. Such an asymmetry can certainly be attributed to doping segregation [18]. The retained value is -1V.

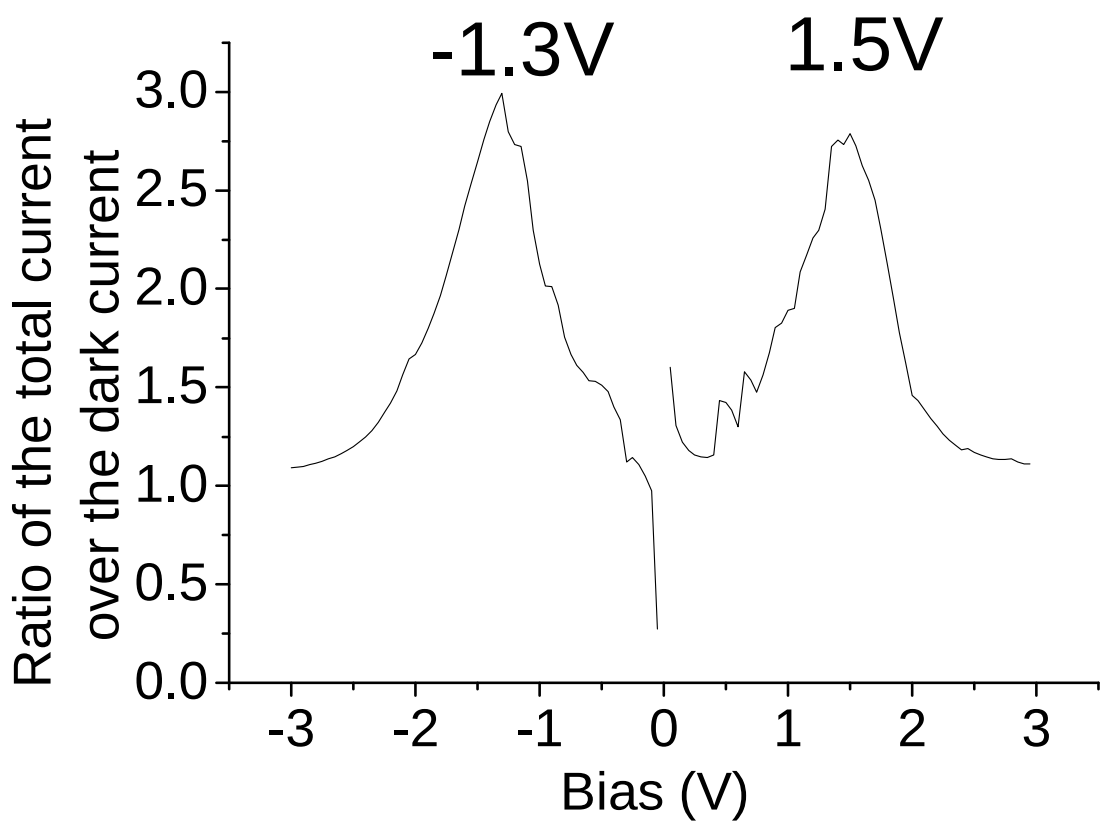

FIG. 5 Ratio of the total current to the dark current as a function of the applied bias.

In the following we present the result of the measurements which have been used as input parameter for the simulation program. Dark current and differential resistance ($R_d$) are presented as a function of the detector temperature, see FIG. 6. The dark current exhibits two regimes. Below 30K the current is almost constant (tunnelling regime), whereas above 30K the current rises with the temperature meaning that the transport is in its thermionic regime. On the contrary the differential resistance starts to decrease when the temperature exceeds 20K. The noise gain and the external quantum efficiency are almost constant with the temperature and respectively equal to 1.1 and 7.1%.

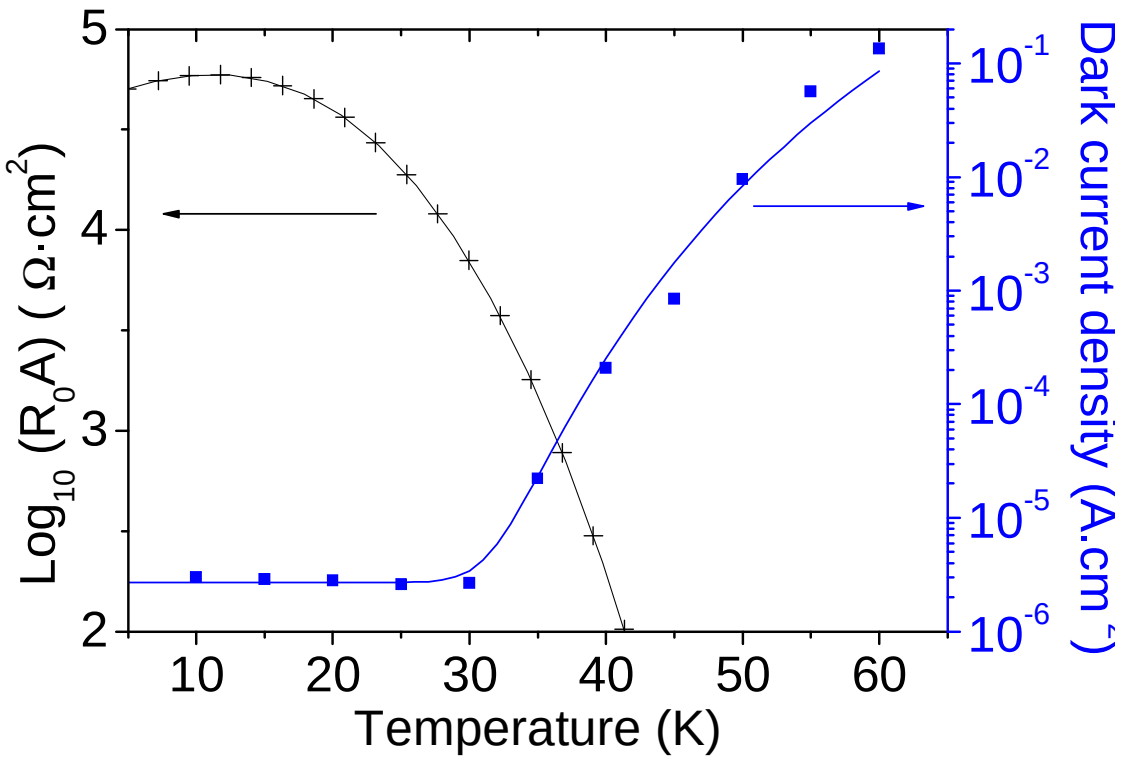

FIG. 6 Dark current and $R_dA$ measurements as a function of the detector temperature for an applied bias of -1V. Dots for experimental points and line for their associated fits.

As said before, we use the code to evaluate the performance of a QWIP based camera. Measurements made on our 15µm QWIP are used as input for the detector part, see FIG. 6. For all the other parts (read-out circuit, matrix format and pitch, integration capacity…) we used realistic components which already exist in the literature or which are commercially available. The tab. II gives the parameters which have been used for the simulation. The integration is adjusted by the code in order that the read out circuit capacity is filled while exposed to a 140K scene. The focal length and pupil diameter have been modified since the Crystal camera to adapt the optical block to the QWIP FPA tab. II Parameters used for the simulation.

| Quantity | Value |
|---|---|
| External read out circuit | 2V |
| Integration Capacity | 0.3pF |
| Noise of the read out circuit | 300µV |
| FPA format | 512×512 |
| Matrix pitch | 25µm |
| Pixel size | 23.5µm |
| Cut-off wavelength of the optical part | 20µm |
| Global optical transmission of the system | 25% |
| Focal length | 15mm |
| Pupil diameter | 6mm |

### 4.3. Results

The results of the simulation are given in FIG. 7. For a detector temperature under 30K, the NETD (noise equivalent temperature difference) is almost constant and equal to 130mK, see FIG. 7 (a). Consequently the QWIP succeed in imaging this 120K scene. Thus the operating temperature can be increased of 20K compared to the Si:Ga FPA. Above 30K the NETD rises. The limit NETD required for this scenario of 200mK is reached for a temperature of 32K. This rise of the NETD goes with a decrease of the integration time.

Physically the limit of 30K corresponds to the transition of the dark current between the tunnel regime and the thermionic regime. Thus the required level of performance implies to use the detector in its tunnel regime.

In all the range of temperature the main noise is the one linked to the dark current, see FIG. 7 (b). We never reach the BLIP regime, even if the signal to noise ratio is sufficient to fulfil the mission. Thus we highlight that even for low flux detection the BLIP regime is not required. Above 30 K the dark current rises and thus the program adapts the integration times to prevent a saturation of the capacity. This decrease goes with a reduced Johnson and photon noise.

We plot the filling factor of the capacity as a function of the temperature of the detector. We can see on FIG. 7 (c) that at least one half of the capacity if filled by the dark current. The useful signal represents less than 10% of the capacity filling. Above 30K the full linearity range results in a complete filling of the capacity, and thus involves a decreasing integration time.

Clearly the performance of this detector can be improved by reducing the dark current, since it will allow a more important part for the useful signal in the capacity and a reduced noise.

Thanks to its important FPA size (512×512) the QWIP can perform simultaneously the imaging of the whole model (large field) with a high resolution (resolution is around 1.6mm on the model). Thus it is no longer useful to use a double stage of optical device. It will also permit an increase of the image quality thanks to the high uniformity of the GaAs matrix.

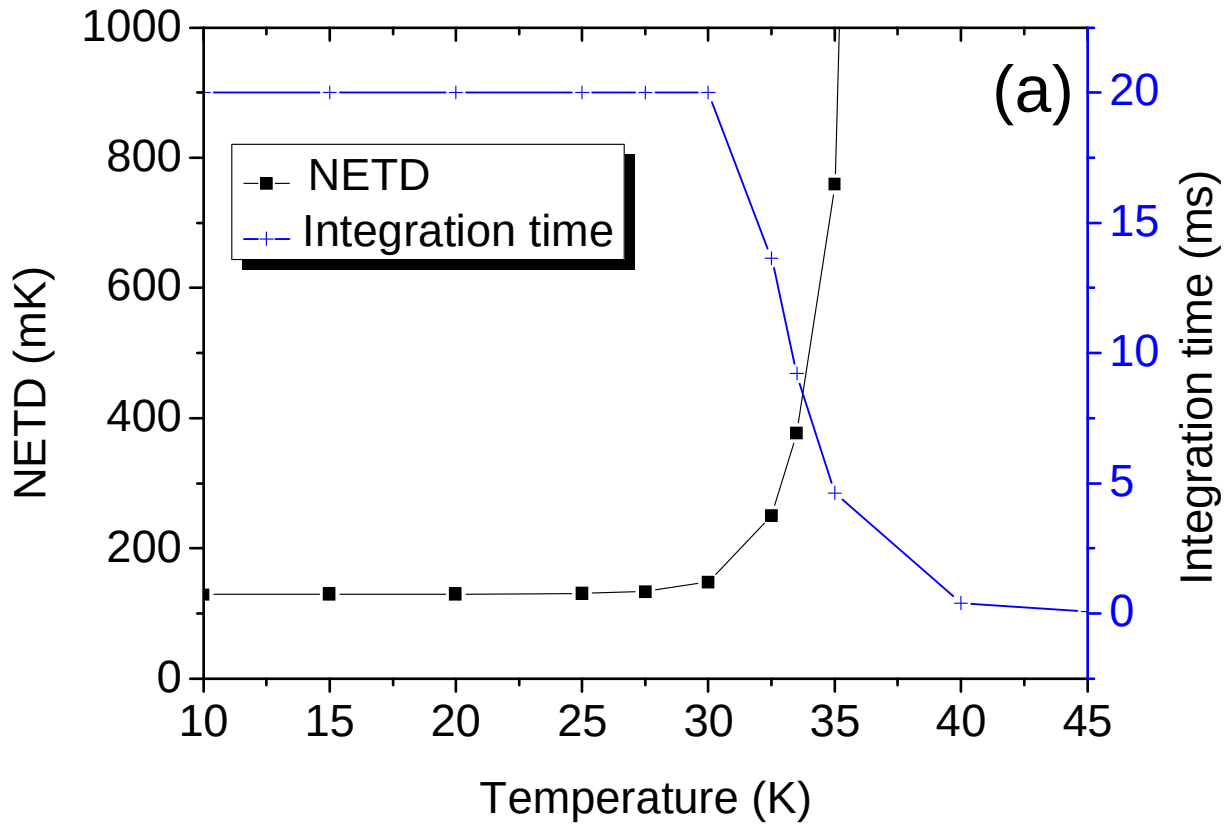

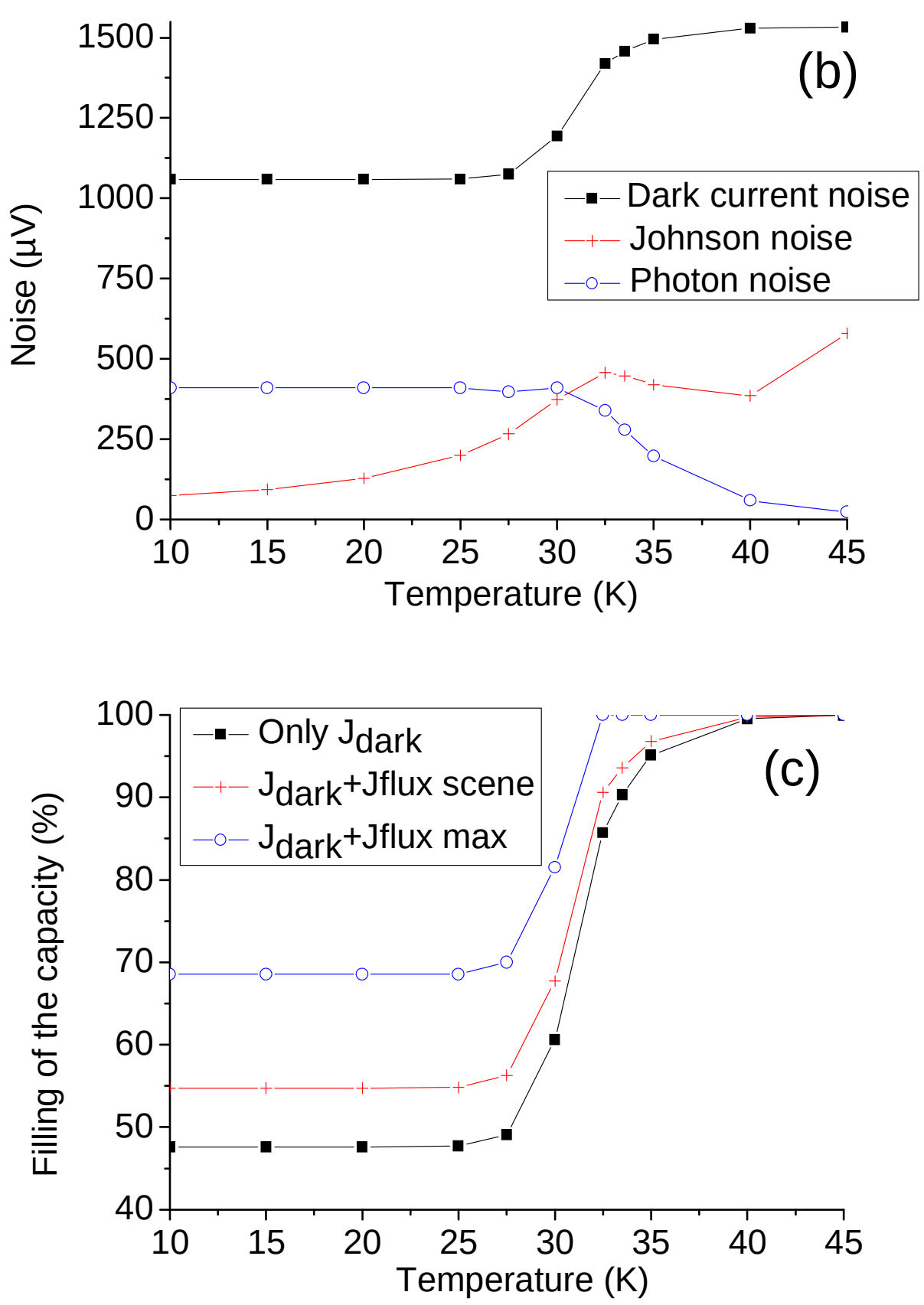

FIG. 7 (a) NETD and integration time as a function of the detector temperature. (b) Different sources of noise as a function of the detector temperature. (c) Filling factor of the read-out capacity as a function of the detector temperature.

## 5. Conclusion

To conclude we have presented a full characterization of a 15µm QWIP illuminated under a low infrared flux. We predict that the exchange of the current Si:Ga camera of the ETW cryogenic wind tunnel can be considered. QWIP succeeds in imaging a scene with a temperature as low as 120K. QWIP would allow an increase of 20K of the operating temperature compared to the current detector and a great simplification of the optical block. The dark current is limiting the level of performance of this component. Its reduction in the tunnel regime is under investigation.

## 6. Acknowledgements

The authors gratefully acknowledge Jérome Primot for many helpful discussions.